# Controllable thickness inhomogeneity and Berry-curvature-engineering of anomalous Hall effect in SrRuO$_3$ ultrathin films

**Authors**

Lingfei Wang[1,2]*, Qiyuan Feng[3,4], Han Gyeol Lee[1,2], Eun Kyo Ko[1,2], Qingyou Lu[3,4]*, and Tae Won Noh[1,2]*

**Affiliations**

[1]Center for Correlated Electron Systems (CCES), Institute for Basic Science (IBS), Seoul, Republic of Korea

[2]Department of Physics and Astronomy, Seoul National University, Seoul, Republic of Korea

[3]Anhui Province Key Laboratory of Condensed Matter Physics at Extreme Conditions, High Magnetic Field Laboratory, Chinese Academy of Sciences, Hefei, Anhui, China.

[4]Hefei National Laboratory for Physical Sciences at the Microscale, University of Science and Technology of China, Hefei, Anhui, China.

Correspondence and requests for materials should be addressed to:

L.F.W. (email: lingfei.wang@outlook.com), Q.Y.L (email: qxl@ustc.edu.cn), and T.W.N. (email: twnoh@snu.ac.kr).

**Abstract**

In quantum matters hosting electron-electron correlation and spin-orbit coupling, spatial inhomogeneities, arising from competing ground states, can be essential for determining and understanding topological properties. A prominent example is Hall anomalies observed in SrRuO$_3$ films, which were interpreted in terms of either magnetic skyrmion-induced topological Hall effect (THE) or inhomogeneous anomalous Hall effect (AHE). To clarify this ambiguity, we systematically investigated the AHE of SrRuO$_3$ ultrathin films with controllable inhomogeneities in film thickness ($t_{SRO}$). By harnessing the step-flow growth of SrRuO$_3$ films, we induced microscopically-ordered stripes with one-unit-cell differences in $t_{SRO}$. The resultant spatial distribution of momentum-space Berry curvatures enables a two-channel AHE, which shows hump-like anomalies similar to the THE and can be continuously engineered via sub-unit-cell control of $t_{SRO}$. In these inhomogeneous SRO films, we microscopically identified a two-step magnetic switching and stripe-like ferromagnetic domains. These features are fingerprints for distinguishing the two-channel AHE from the skyrmion-induced THE.

# Introduction

Interplay between electron-electron (el-el) correlation and spin-orbit coupling (SOC) in quantum matters can stimulate numerous intriguing correlated topological phases(*1*, *2*), such as spin-orbital-coupled Mott insulator(*3*, *4*), Weyl semimetal(*5*, *6*), and Dirac semimetal(*7*, *8*). These exotic phases not only provide a fertile ground for discovering exotic topological physics but also hold great potential for electronic and spintronic devices with high tunability(*9*). In electronic systems with non-trivial topology, except for the topologically-protected edge/surface states, the physical properties and band structures are mostly presumed to be homogenous(*10*, *11*). Nevertheless, in strongly-correlated electronic systems, it is well-known that coupling between spin, charge, lattice, and orbital degrees of freedom facilitates competing ground states and thus microscopic phase segregation(*12*, *13*). Recent works on quantum anomalous Hall systems and non-collinear antiferromagnetic insulators further indicate that spatial inhomogeneities, such as magnetic domain structures and domain walls, can be important in understanding the physical properties of topological systems with considerable *el-el* correlation(*14*, *15*). However, the lack of controllability hinders systematic investigation on the essential roles of spatial inhomogeneity in correlated topological systems.

Magnetotransport properties and Berry curvature are two powerful tools for understanding topological physics in magnetic systems(*9*, *16*–*19*). Berry curvature in three-dimensional momentum-space (*k*-space) is an intrinsic topological property of a given electronic band structure(*20*). In magnetic systems with broken time-reversal symmetry, the intrinsic anomalous Hall effect (AHE) is determined by the integral of the *k*-space Berry curvature over the Brillouin zone(*18*, *20*, *21*). Meanwhile, a spin-polarized itinerant electron can pick up an additional Berry phase when passing through a chiral magnetic structure(*22*). In this case, non-zero Berry curvature acts as a real-space fictitious magnetic field (*H*) and gives rise to the topological Hall effect (THE). This Hall signal, distinct from the AHE, has been observed in a variety of magnetic systems hosting magnetic skyrmions and other types of non-collinear spin ordering(*23*–*25*).

Non-vanishing real- and *k*-space Berry curvatures can coexist in one magnetic system. For example, in antiferromagnetic Heusler compound $Mn_3X$ (X = Ga, Ge or Sn), *k*-space Berry curvature arises from the Weyl points in band structures, leading to a large AHE comparable to typical ferromagnets(*19*, *26*, *27*). Moreover, the scalar spin chirality also gives rise to a real-space Berry curvature and the THE(*28*). The entanglement between real- and *k*-space Berry curvatures may bring challenges in interpreting the magnetotransport results. In doped EuTiO$_3$, a magnetic semiconductor, non-monotonic Hall signal was observed during the magnetization process. This Hall anomaly was ascribed to either magnetic skyrmions (i.e., non-zero real-space Berry curvature) or Zeeman splitting-modulated band structure (i.e., non-zero k-space Berry curvature)(*29*, *30*). It is expected that the presence of microscopic spatial inhomogeneity would make the entanglement between Berry curvature and magnetotransport more complicated.

Perovskite-structured oxide SrRuO$_3$ (SRO) is a 4d itinerant ferromagnet(*31*), in which both real- and k-space Berry curvatures are critical for determining and understanding its intriguing topological properties. In SRO epitaxial heterostructures hosting a fine balance between *el-el* correlation and SOC, interfacial inversion-symmetry-breaking can activate the Dzyaloshinskii-Moriya interaction (DMI). As a consequence, magnetic skyrmions and chiral magnetic domains can be stabilized and embedded in a ferromagnetic matrix(*32–36*). The resultant inhomogeneity in real-space Berry curvature inevitably induces a superposition of THE and AHE signals (Fig. 1, A and B), thus giving rise to hump-like anomalies in the *H*-dependent Hall resistance ($R_{xy}$-*H*) curves (Fig. 1E). This Hall signal was first reported by J. Matsuno *et al.* in SRO/SrIrO$_3$ bilayers and then discovered in numerous SRO-based heterostructures(*32–35*). Soon after the discoveries of THE in SRO heterostructures, an alternative interpretation based on spatial inhomogeneity in AHE and k-space Berry curvature was proposed(*37–39*). The Ru $t_{2g}$ bands consist of nearly degenerated points near the Fermi surface ($E_F$) (*39*, *40*). The so-called "k-space magnetic monopoles" makes the integral Berry curvature and AHE very sensitive to various tuning factors including temperature (*T*), band structure, $E_F$ position, chemical doping, magnetism, and structure distortions(*40*, *41*). A spatial inhomogeneity in these tuning factors, if exists in SRO films, may result in two types of AHE signals with a reversed sign and unequal coercive fields ($H_C$) (Fig. 1, C and D)(*37*, *39*). The superposition of these two AHE loops can also generate hump-like $R_{xy}$ anomalies (Fig. 1E). Namely, the possible spatial inhomogeneity in either real- or k-space Berry curvatures renders a clear interpretation of Hall signals in SRO. For clarifying this confusion and transferring the topological properties of SRO to practical electronic/spintronic devices, experimental routes to effectively control and exclusively distinguish these spatial inhomogeneities are highly desirable.

Motivated by this unsolved controversy, here we systematically investigated the interplay between AHE and a controllable thickness inhomogeneity in ultrathin SRO films. By harnessing the step-flow growth of SRO films, we were able to induce a microscopic inhomogeneity in SRO film thickness ($t_{SRO}$). The regions with one-unit-cell $t_{SRO}$ differences were self-organized into well-aligned stripe patterns. Owing to the strong coupling between $t_{SRO}$, ferromagnetism, and k-space Berry curvature, the artificially-induced thickness inhomogeneity gives rise to two sets of independent and microscopically-ordered AHE channels. In this two-channel AHE model, the cumulative $R_{xy}$ signal shows $H$-dependent anomalies similar to the THE and can be continuously tuned via nominal $t_{SRO}$. Thanks to the highly-ordered characteristic of thickness inhomogeneity, we further identified the associated two-step magnetic switching behavior and stripe-like ferromagnetic domains microscopically. These two features are critical fingerprints for exclusively distinguishing SRO-based heterostructures with the two-channel AHE from the ones with the skyrmion-induced THE.

## Results

### Control of SRO films thickness and artificial thickness inhomogeneity

SRO thin films were epitaxially grown on $TiO_2$-terminated $SrTiO_3(001)$ [STO(001)] substrates via pulsed laser deposition(*34*). The substrate temperature and laser repetition rate were optimized and set to 700 °C and 2 Hz, respectively (See procedural details in the Materials and Methods section). The film growth was monitored *in-situ* via reflection high-energy electron diffraction (RHEED). The time-dependent RHEED intensity profiles of the specular point are shown in Fig. 2A. The RHEED intensity reaches its first maximum after the initial growth of three atomic layers (SrO-$RuO_2$-SrO), denoted as a 1.5 unit-cells (u.c.) thick layer. This RHEED oscillation, as previously reported(*42*), signifies a surface termination conversion from $RuO_2$ to SrO. The RHEED intensity reaches a second maximum after another 1 u.c. growth and then saturates, signifying a growth mode transition from layer-by-layer to step-flow(*43*). The growth rate ($N_{uc}$ laser pulses per u.c.) was calculated according to these two initial RHEED oscillations, and a sub-unit-cell control of $t_{SRO}$ can be achieved by setting the total number of laser pulses ($N_{total}$). As schematically shown in Fig. 2B, the step-flow growth always starts from the one-unit-cell-high terrace edges. By terminating the deposition before one SRO monolayer covering the entire surface, we can grow an SRO film consisting of two sets of alternatively aligned stripe regions with one-unit-cell-difference in $t_{SRO}$. We define the nominal SRO film thickness $t_{SRO} = (N_{total} - 0.5N_{uc})/N_{uc}$. According to this definition, an integer $t_{SRO}$ value corresponds to an SRO film with a uniform thickness, while a non-integer $t_{SRO}$ corresponds to a film with artificial thickness inhomogeneity. For instance, in an SRO film with nominal $t_{SRO}$ = 4.3 u.c., the 4.0 (5.0) u.c.-thick area is 70% (30%).

To systematically investigate the evolution of magnetotransport properties of SRO films with $t_{SRO}$, we grew a series of films with integer $t_{SRO}$ ranging from 3.0 to 10.0 u.c. and another series of films with non-integer $t_{SRO}$ ranging from 4.1 to 4.9 u.c. The temperature-dependent longitudinal transport and magnetism were characterized (fig. S1 and fig. S2). All of the films show a systematic decay in ferromagnetism and metalicity as $t_{SRO}$ decreases (*31–34*). We also grew a comparison SRO film with a nominal $t_{SRO}$ = 4.0 u.c. using a high laser repletion rate of 10 Hz. The high deposition flux causes disturbed step-flow growth mixed with two-dimensional island formation(*44*), as implied by the gradually decreasing RHEED intensity (Fig. 2A).

**AHE sign reversal in SRO ultrathin films**

We first characterized the transverse magnetotransport of SRO films with integer $t_{SRO}$. The Hall resistance of SRO can be expressed as $R_{xy} = R_0 H + R_{AHE}$. The first term corresponds to the ordinary Hall effect, where $R_0$ is the ordinary Hall coefficient. For clarity, we subtracted this term by linear fitting the $R_{xy}$-$H$ curves in the range of $\mu_0 H \geq 3.0$ T. The second term, the anomalous Hall resistance, can be described by $R_{AHE} = R_S M$, where $R_S$ and $M$ correspond to the cumulative anomalous Hall coefficient and out-of-plane magnetization, respectively(*18, 31*). The $R_{AHE}$-$H$ curves measured from SRO films with various integer $t_{SRO}$ are shown in Fig. 3, A to D. At 10 K, the $R_{AHE}$-$H$ curves of 10.0 u.c. SRO films show a reversed magnetic hysteresis loop (Fig. 3A), signifying a negative $R_S$. Such a bulk-like AHE persists as $t_{SRO}$ reduces to 5.0 u.c. (Fig. 3B). As $t_{SRO}$ decreases further to 4.0 u.c. (Fig. 3C), the $R_{AHE}$-$H$ curves start to behave like normal magnetic hysteresis loops, indicating a dramatic sign reversal in $R_S$ from negative to positive. The positive $R_S$ remains as $t_{SRO}$ reduces further to 3.0 u.c. (Fig. 3D). We also measured the $R_{AHE}$-$H$ curves at various temperatures (*T*). For the films with $t_{SRO} \geq 5.0$ u.c., a sign crossover in AHE can be also observed as *T* increases up to ~120 K. In contrast, for the films with $t_{SRO} <$ 5.0 u.c., the sign of $R_S$ remains positive over the entire *T* range below the Curie temperature ($T_C$). Notably, all of the $R_{AHE}$-$H$ curves measured from SRO single layer films with integer $t_{SRO}$ do not show any signs of the THE, possibly due to the negligible inversion symmetry breaking and DMI (fig. S3).

AHE in SRO is dominated by the intrinsic origin, *i.e.*, the k-space Berry curvature. Therefore, its sign reversal is closely related to the changes in electronic band structures. *T*-driven AHE sign crossover has been widely observed and studied in SRO films and single crystals. This behavior can be understood by modulation of integrated Berry curvature as $E_F$ shifts around the avoided band-crossing points in Ru $t_{2g}$ bands during *T* scanning(*31, 39, 40*). Based on this scenario, we suggest that the AHE sign reversal as $t_{SRO}$ decreases from 5.0 to 4.0 u.c. should also have a $k$-space Berry curvature-related origin. As $t_{SRO}$ reduces, possible enhancement in spatial confinement effect, charge depletion, *el-el* correlation, and structural distortions could induce sizable modulations in both $E_F$ and band structure(*45–47*). The associated changes in the integral Berry curvature over the Brillouin zone may induce the $t_{SRO}$-driven AHE sign reversal. Detailed evolutions of band structure and k-space Berry curvature still require in-depth studies via angle-resolved photoemission spectroscopy and density functional theory simulations.

**Tunable AHE in SRO films with inhomogeneous $t_{SRO}$**

Despite the elusive mechanism, the $t_{SRO}$-driven AHE sign reversal between the 4.0 and 5.0 u.c. SRO films (Fig. 4, A and G) offers a effective modulation of AHE via artificial thickness inhomogeneity. $R_{AHE}$-$H$ curves (at 10 K) of SRO films with non-integer nominal $t_{SRO}$ ranging from 4.1 to 4.9 u.c. are shown in Fig. 4, B to F. In these films with inhomogeneous $t_{SRO}$, the $R_{AHE}$-$H$ curves exhibit two important features as the nominal $t_{SRO}$ decreases. First, the saturated $R_{AHE}$ at 5.0 T (-5.0 T) shows a clear sign crossover from negative to positive (positive to negative). Second, $R_{AHE}$ near $H_C$ shows a hump-like feature, similar to the previously reported THE signals. This hump-like anomaly reaches the maximum amplitude at $t_{SRO}$ = 4.5 u.c., which gives an identical area of the 5.0 and 4.0 u.c.-thick regions. Both features can be well reproduced by linear superposition of the $R_{AHE}$-$H$ curves measured from the 4.0 and 5.0 u.c. SRO films (light grey curves in Fig. 4, B to F). These results strongly suggest that AHE in these inhomogeneous SRO films is dominated by two sets of independent magnetotransport channels with distinct $k$-space Berry curvatures. Taking the $t_{SRO}$ = 4.5 u.c. film as an example, such a two-channel AHE is schematically depicted in Fig. 4, J to L. At a saturated $H$ value ($\mu_0 H$ = 2.0 T), $M$ in the 4.0 and 5.0 u.c. channels align parallel, and the $R_{AHE}$ signals from these two channels cancel out due to their opposite signs (Fig. 4, J and K). As a consequence, the overall $R_{AHE}$ value is close to zero. In contrast, at a critical $H$ value between the $H_C$ of the 5.0 and 4.0 u.c. channels ($\mu_0 H$ = -1.0 T), $M$ in the two channels align antiparallel and the $R_{AHE}$ signals share the same sign, which should accumulatively contribute to the overall $R_{AHE}$ and thus induce the hump-like anomalies (Fig. 4, J and L). For the films grown in optimized condition, the surface topography shows well-defined one-unit-cell-high terrace structures with uniform width and straight edges (Fig. 4I, top panel). In this case, the stripe-shaped AHE channels are well-aligned with the terrace edges and the widths can be continuously modulated by varying nominal $t_{SRO}$. Therefore, these SRO ultrathin films with artificial thickness inhomogeneity can enable highly-ordered and controllable magnetotransport for spintronic applications(*48*).

Disorder in the terrace structure also plays an important role in determining the two-channel AHE. For the 4.0 u.c.-thick SRO film grown with the high laser repetition rate of 10 Hz, the disturbed step-flow growth leads to curved terrace edges and numerous one-unit-cell-high islands at the surface (Fig. 4I, bottom panel). As schematically depicted in the bottom panel of Fig. 2B, such a highly-disordered surface topography inevitably leads to significant inhomogeneity in $t_{SRO}$. The corresponding $R_{AHE}$-$H$ curve (Fig. 4H) shows obvious humps peaked ~ ±0.8 T and gradually increased slopes at the higher $H$, which should be assigned to a mixture of $R_{AHE}$ signals in randomly distributed 3.0, 4.0, and 5.0 u.c. regimes. This strongly suggests that spatial inhomogeneities, as well as the AHE in ultrathin SRO films, are extremely sensitive to the growth condition. When the growth conditions deviate from the optimized window, unexpected spatial inhomogeneity in the band structure and ferromagnetism could occur due to disordered terraces, clusters with Ru-deficiency, structure domains, and imperfect interfacial/surface regions. The associated inhomogeneity in k-space Berry curvature can be intricate and inevitably induces hump-like signals in $R_{AHE}$. Minimizing these artifacts in AHE signal would be an essential prerequisite for investigating the THE in SRO-based heterostructures.

**Identifying the magnetic inhomogeneity in SRO films**

So far, we have found that inhomogeneities in both real-space Berry curvature (from magnetic skyrmions) and *k*-space Berry curvature (from band-structure modulations) can induce hump-like anomalies in $R_{AHE}$-$H$ curves. One important question pertains to how we can distinguish these two scenarios experimentally. According to Fig. 3 and fig. S1, $t_{SRO}$ is strongly coupled with both the k-space Berry curvature and ferromagnetism. Therefore, in the films with inhomogeneous $t_{SRO}$, the 4.0 and 5.0 u.c. regimes should show clear differences in both *M* and $H_C$. Such spatial magnetic inhomogeneities should have a similar length scale to the terrace structure (several hundreds of nanometers or micrometers). On the contrary, for SRO-based heterostructures with magnetic skyrmions, the uniform $t_{SRO}$ and interfacial DMI(*32*, *34*) should minimize the magnetic inhomogeneities. Therefore, sub-micrometer-scale magnetic inhomogeneity should be a unique characteristic of the SRO films with inhomogeneous $t_{SRO}$, and thus can be utilized for distinguishing the two-channel AHE from the skyrmion-induced THE.

We first identified two-step magnetic switching behavior of the thickness-inhomogeneous films via simple *M-H* characterizations. As shown in Fig. 5A, *M-H* curves of the 4.0 and 5.0 u.c. SRO films at 10 K show similar continuous magnetic switching behaviors but distinct $H_C$: 0.9 and 1.2 T for the 4.0 and 5.0 u.c. films, respectively. As schematically depicted in Fig. 5B, the inhomogeneities in $t_{SRO}$ should not only induce hump-like anomalies in $R_{AHE}$-$H$ curves but also a two-step feature in the $M$-$H$ curves. In the $M$-$H$ curves measured from the 4.3 and 4.5 u.c. SRO films (Fig. 5C), a plateau can be clearly identified between the two successive magnetic switchings. As marked by the dashed lines, the two $H_C$ are close to those of the 4.0 and 5.0 u.c. SRO films. Notably, such a clear two-step feature is rarely observed in the SRO heterostructures with uniform DMI (*32–36*). In these systems, as the $H$ direction reverses, skyrmions or chiral magnetic bubbles emerge from the ferromagnetic matrix and broaden the magnetic switching, leading to gradual slopes rather than two-step features in the $M$-$H$ curves (fig. S4). Therefore, the two-step feature in $M$-$H$ curves can serve as a marker of thickness inhomogeneity as well as the two-channel AHE.

We also used magnetic force microscopy (MFM) to achieve real-space imaging of the microscopic inhomogeneities in ferromagnetism (see Materials and Methods sections for details). All of the MFM images were measured from the 4.5 u.c. SRO film at a pre-selected region (2.5 × 2.5 μm$^2$) with a uniform terrace structure (Fig. 6A). After zero-field cooling the 4.5 u.c. SRO film down to 10 K, the MFM image (Fig. 6B) clearly shows two regions with positive and negative frequency shifts ($\Delta f$), corresponding to ferromagnetic domains with magnetic moment pointing upward and downward, respectively(*35*). By increasing $H$ up to 5.0 T, magnetic moments are uniformly pointing upward, while the MFM image still shows a stripe-like magnetic contrast (Fig. 6C). As marked by the dashed lines, these stripes are well aligned with terrace edges. The line profiles from the topographic and MFM images (Fig. 6D) further confirmed the correspondence between the terrace structure and stripe-like MFM signals. The regions with larger (smaller) $\Delta f$ should be assigned to the 5.0 u.c. (4.0 u.c.) regions with larger (smaller) $M$.

MFM can microscopically identify the two-step magnetic switching behavior of the 4.5 u.c. SRO film. We measured a series of MFM images as $H$ scans from -5.0 to 5.0 T (Fig. 6, E to L). The $H$ range for acquiring MFM images covers the two successive magnetic switchings (Fig. 6O). After scanning $H$ from 5.0 T to -0.2 T, magnetic moment over the entire film is uniformly pointing upward. Thus, the MFM image (Fig. 6E) shows positive $\Delta f$ signal with thickness inhomogeneity induced stripe patterns only. As $H$ increases to -0.6 T, magnetic moments of several magnetic domains in 4.0 u.c.-thick stripe regions start to switch downward, resulting in the negative $\Delta f$ signal in Fig. 6F. Such local magnetic switching behavior becomes more clear as $H$ reaches -0.8 and -1.0 T (Fig. 6, G and H). At -1.0 T, most of the 4.0 u.c.-thick stripe domains are downward polarized (negative $\Delta f$), while the 5.0 u.c.-thick domains remain upward polarized (positive Δ*f*). At -1.1 and -1.2 T (Fig. 6, I and J), the downward-polarized domains start to expand into the 5.0 u.c.-thick regions, leading to the second magnetic switching. This magnetic switching is sharper than the first one occurring in the 4.0 u.c.-thick regions. Consequently, all the magnetic domains completely switch to downward-polarized at -1.3 T, and the MFM images at -1.3 and -1.5 T show weak and stripe-like contrasts from thickness inhomogeneity only. The evolution of magnetic domain configurations and corresponding Δ*f* profiles during the above *H* scanning process are schematically depicted in Fig. 6P.

To further clarify the magnetic domain switching behavior, we applied a pixel-by-pixel subtraction operation to the adjacent MFM images acquired at -1.1 to -1.3 T(*34*), at which the $R_{AHE}$ humps appear. As shown in Fig. 6, M and N, the subtracted MFM images show magnetic domains on the scale of several hundred nanometers or even micrometers, aligned and elongated along the terrace edges. These magnetic domain patterns are consistent with the two successive sharp magnetic switchings observed in *M-H* curves: As *H* scans over $H_C$, the magnetic moments in micrometer-sized domains flips collectively. In these SRO films with strong magnetic anisotropy and negligible DMI, such sharp magnetic switching can help to reduce the energy cost from ferromagnetic domain walls. By contrast, in the SRO heterostructures with sizable DMI, magnetic switchings are always accompanied by the emergence of magnetic skyrmions, magnetic bubble domains with chirality, and Neel-type domain walls. The corresponding MFM contrasts usually have circular or curved shapes, and the typical length scale is within several tens of nanometers(*24*, *34*, *35*). On this basis, our MFM results provide critical experimental evidence for distinguishing the SRO films with thickness inhomogeneity and two-channel AHE from those with magnetic skyrmions and THE.

## Discussion

We systematically investigated the evolution of AHE with artificial thickness inhomogeneity in ultrathin SRO films. By harnessing the step-flow growth of SRO films, we induced a microscopically-ordered thickness inhomogeneity in a controlled manner. The sizable *el-el* correlation and non-trivial band topology enable a strong coupling between $t_{SRO}$ and k-space Berry curvature in these SRO ultrathin films. Consequently, the artificially-induced thickness inhomogeneity enables an effective Berry-curvature-engineering of AHE. The cumulative signal from two sets of highly-ordered AHE channels shows hump-like anomalies, which were previously assigned to the skyrmion-induced THE. Owing to the microscopically-ordering nature and high controllability of this artificially-induced thickness inhomogeneity, we can identify the associated microscopic inhomogeneity in ferromagnetism through *M-H* measurements and MFM. Both experimental characterizations can be seen as explicit evidence for differentiating SRO films with two-channel AHE from those with intrinsic THE.

This work provides new insights into the role of spatial inhomogeneity in correlated topological phases. The comprehensive characterizations on the AHE and microscopic inhomogeneities may help to resolve the long-standing challenge associated with interpreting the Hall anomalies of SRO-based heterostructures. The method for inducing and controlling artificial inhomogeneity at nanoscale can be applied to other epitaxial oxide systems. It paves experimental routes to effectively harness the correlated topological phases and transfer their intriguing functionalities for practical device applications.

## Materials and Methods

**Sample preparation.** SRO thin film was deposited onto STO(001) substrates using pulsed laser deposition. The as-received STO(001) substrates (Crystec) were treated with buffered hydrofluoric acid and annealed in the ambient atmosphere to achieve an atomically-flat and $TiO_2$-terminated surface with one-unit-cell-high terraces. The SRO ultrathin films were grown under an oxygen partial pressure of 100 mTorr. The substrate temperature was maintained at 700 ℃ during the deposition. SRO polycrystalline target was ablated by a KrF excimer laser ($\lambda$ = 248 nm; Coherent) with a laser fluence of ~2 J/cm$^2$. The laser repetition rate was set at 2 Hz, leading to an optimal growth rate of ~60 pulses per u.c ($N_{uc}$). The nominal $t_{SRO}$ was controlled precisely by *in-situ* RHEED monitoring and the number of laser pulses ($N_{total}$). For SRO films with nominal $t_{SRO}$ = 4.0, 4.1, 4.3, 4.5, 4.7, 4.9 and 5.0 u.c., the corresponding $N_{total}$ = 270, 277, 288, 300, 311, 325, and 333 pulses.

**Low-temperature electrical transport and magnetization measurements** The longitudinal and transverse transport data at low temperature were measured using a Physical Properties Measurement System (PPMS, Quantum Design) on standard Hall bars. Conventional photolithography and ion-milling were used to pattern the SRO films into the Hall bar geometry. The channel size was 50 × 50 μm$^2$ and the current flow direction was aligned with the terrace edges. After patterning, the samples were ex-situ annealed at 600 °C in ambient oxygen flow for 1 h to minimize the oxygen deficiency-induced during growth and ion-milling. Ti (5 nm) and Pt (50 nm) films were sputtered onto the Hall bar as contact electrodes. *M-H* curves at 10 K were measured using a SQUID magnetometer (MPMS; Quantum Design) with *H* applied along the out-of-plane direction.

**MFM experiments.** MFM experiments were performed using a custom-designed variable-temperature MFM system, equipped with a 20 T superconducting magnet. We incorporate a commercial piezoresistive cantilever (PRC400; Hitachi High-Tech Science Corporation) as the force sensor. The resonant frequency of the cantilever is about 42 kHz. The MFM tip has a magnetic coating consisting of 5 nm Cr, 50 nm Fe, and 5 nm Au films. This magnetic coating was magnetized perpendicular to the cantilever. The $H_C$ and saturation fields are ~250 Oe and ~2,000 Oe, respectively. A built-in phase-locked loop (R9 controller; RHK Technology) was used for MFM scanning control and signal processing. MFM images were collected in a constant height mode. First, a topographic image was obtained using contact mode, from which the sample surface tilting along the fast and slow scan axes could be compensated. Then the tip was lifted by ~50 nm to the surface and MFM images were obtained in frequency-modulation mode. The resonant frequency shift $\Delta f$ of the MFM cantilever is given by:

$$\frac{\Delta f}{f_0} = -\frac{1}{2k}\frac{\partial F}{\partial z} \propto \frac{\partial^2 B_z}{\partial^2 z^2} \qquad (1)$$

where $f_0$, $F$, and $k$ denote the resonance frequency, the magnetic force between the MFM tip and the sample surface, and the spring constant of the cantilever, respectively. According to equation (1), the resonant frequency shift is only sensitive to the out-of-plane (*z*-direction) component of the stray field ($B_z$). When the magnetic moment of a ferromagnetic domain lies parallel (antiparallel) to the external $H$, the magnetic force between the tip and ferromagnetic domain is attractive (repulsive), leading to a negative $\Delta f$ (blue colored regions in MFM images) [positive $\Delta f$ (red-colored regions in MFM images)]. MFM images were analyzed using Gwyddion software and custom-written MATLAB codes.

## Supplementary Materials

Fig. S1. Longitudinal transport and magnetization of SRO films with various $t_{SRO}$.

Fig. S2. Longitudinal transport of SRO films with inhomogeneous $t_{SRO}$.

Fig. S3. Scanning transmission electron microscopy on SRO/STO interface.

Fig. S4. Magnetic hysteresis loops of BaTiO$_3$/SRO bilayers with magnetic skyrmions.

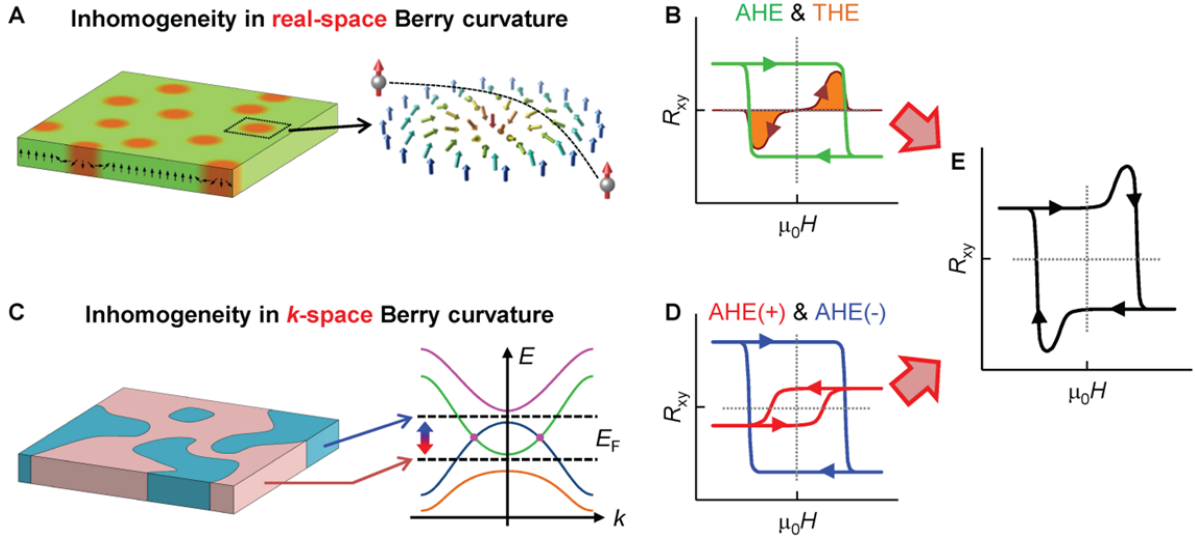

**Fig. 1. Two-channel anomalous Hall effect and magnetic skyrmion-induced topological Hall effect.** (**A**) Schematic illustration of topological Hall effect (THE), which originates from the real-space Berry curvature acquired by the spin-polarized electron when it passes through a magnetic skyrmion. The corresponding magnetic-field-dependent Hall resistance ($R_{xy}$-$H$) curve (**B**) consists of both the anomalous Hall effect (AHE) and THE signals. (**C**) Schematic illustration of inhomogeneity in *k*-space Berry curvature, which originates from the band structure change and Fermi level ($E_F$) shift with respect to the avoided band crossing point. The corresponding $R_{xy}$-$H$ curve (**D**) consists of two AHE channels with distinct hysteresis loops. Both the skyrmion model and two-channel AHE model can induce a hump-like $R_{xy}$ signal shown in (**E**).

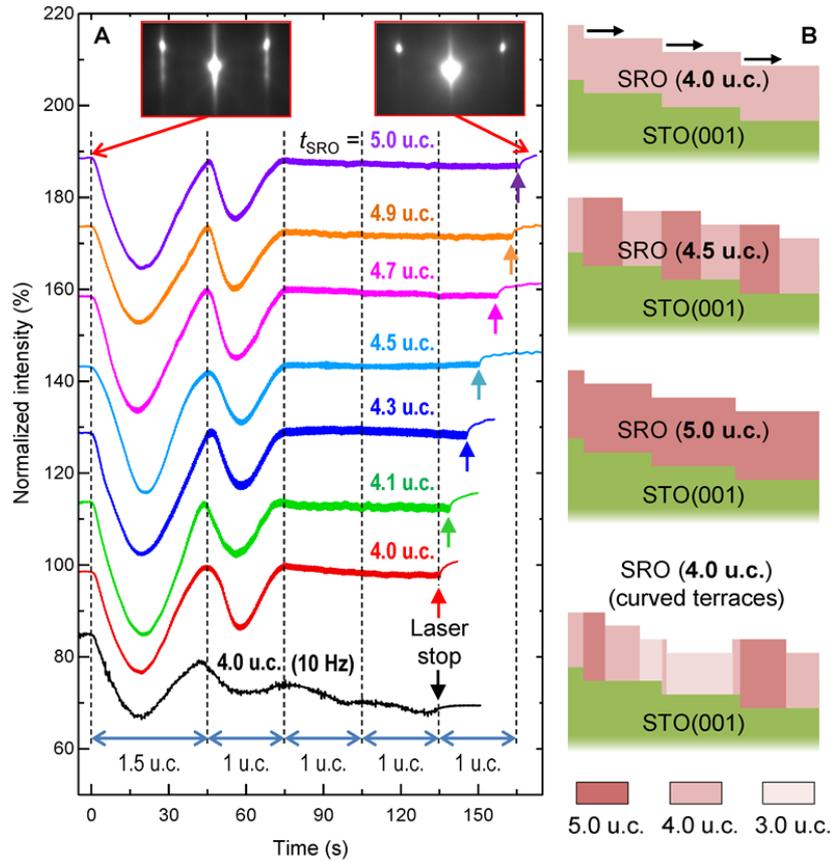

**Fig. 2. Growth of SrRuO₃ films with controlled thickness inhomogeneity.** (**A**) Time-dependent reflection high-energy electron diffraction (RHEED) intensity profile of the specular spot during the growth of SrRuO₃ (SRO) thin films with various nominal thickness $t_{SRO}$. All of the RHEED intensity profiles clearly show two oscillations and then saturate, signifying a surface termination conversion and a growth mode transition from layer-by-layer to step-flow. We calculated the growth rate from the two RHEED oscillations during layer-by-layer growth and precisely controlled $t_{SRO}$ through the number of laser pulses. The time points when laser stops are marked by solid arrows, and the complete growth of each SRO monolayer is marked by dashed lines. The RHEED profile for SRO film growth with a high laser repetition rate of 10 Hz is also inserted for comparison. The decay in RHEED intensity implies a disturbed step-flow growth mode. (**B**) Schematics of SRO film growth following a perfect and disturbed step-flow mode. The step flow directions are marked by solid arrows. The 3.0, 4.0 and 5.0 unit cells (u.c.) -thick SRO regions are marked in light pink, pink, and brown, respectively.

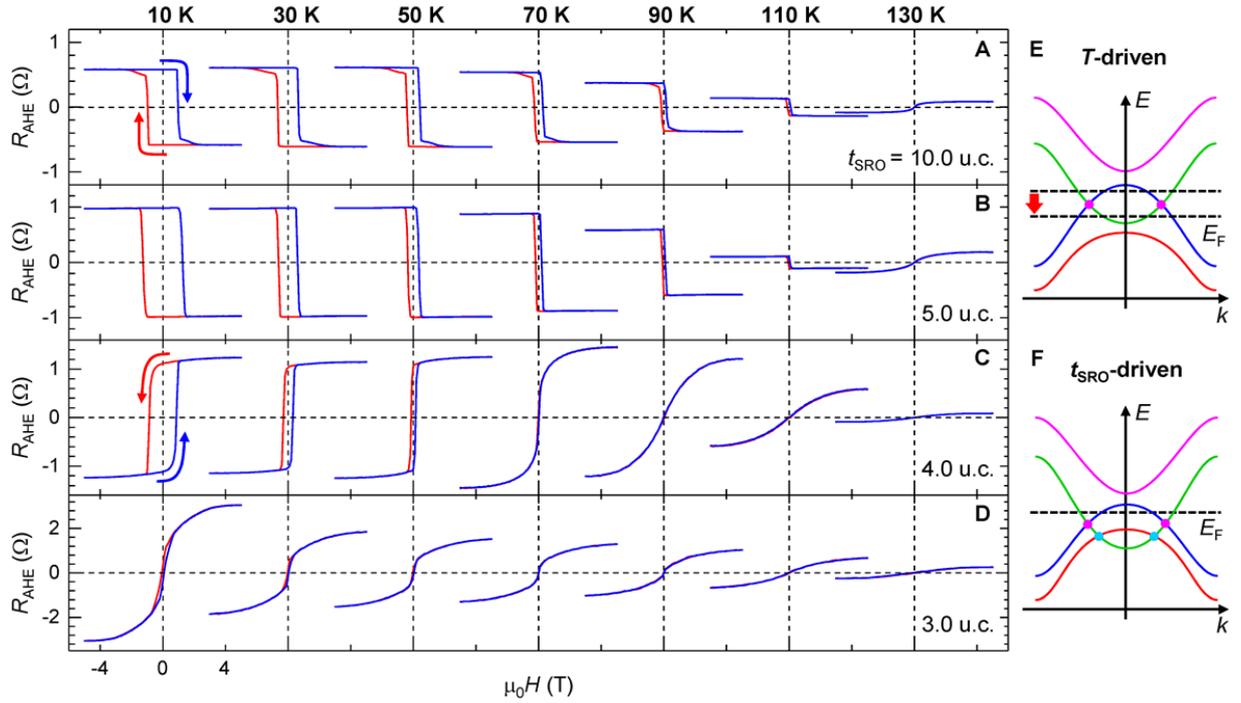

**Fig. 3. Evolution of anomalous Hall effect with various $t_{SRO}$ and temperature.** (**A** to **D**) $H$-dependent Anomalous Hall resistance ($R_{AHE}$-$H$) curves measured from SRO films with various $t_{SRO}$ and temperatures ($T$) ranging from 10 to 130 K. The $H$ scanning direction is marked by solid arrows. At 10 K, the sign of the anomalous Hall coefficient changes from negative to positive as $t_{SRO}$ decreases to 4.0 u.c. For the 10.0 and 5.0 u.c. films, as $T$ increases up to 130 K, $R_S$ sign reversal also occurs. (**E**) Schematics for illustrating the sign change in AHE. For the SRO films with $t_{SRO} \geq 5$ u.c., the $T$-driven sign reversal in AHE can be understood by the Fermi level ($E_F$) shift with respect to the avoided band crossing points, while the $t_{SRO}$-driven sign reversal in AHE could originate from the changes in the band structure and associated $k$-space Berry curvature.

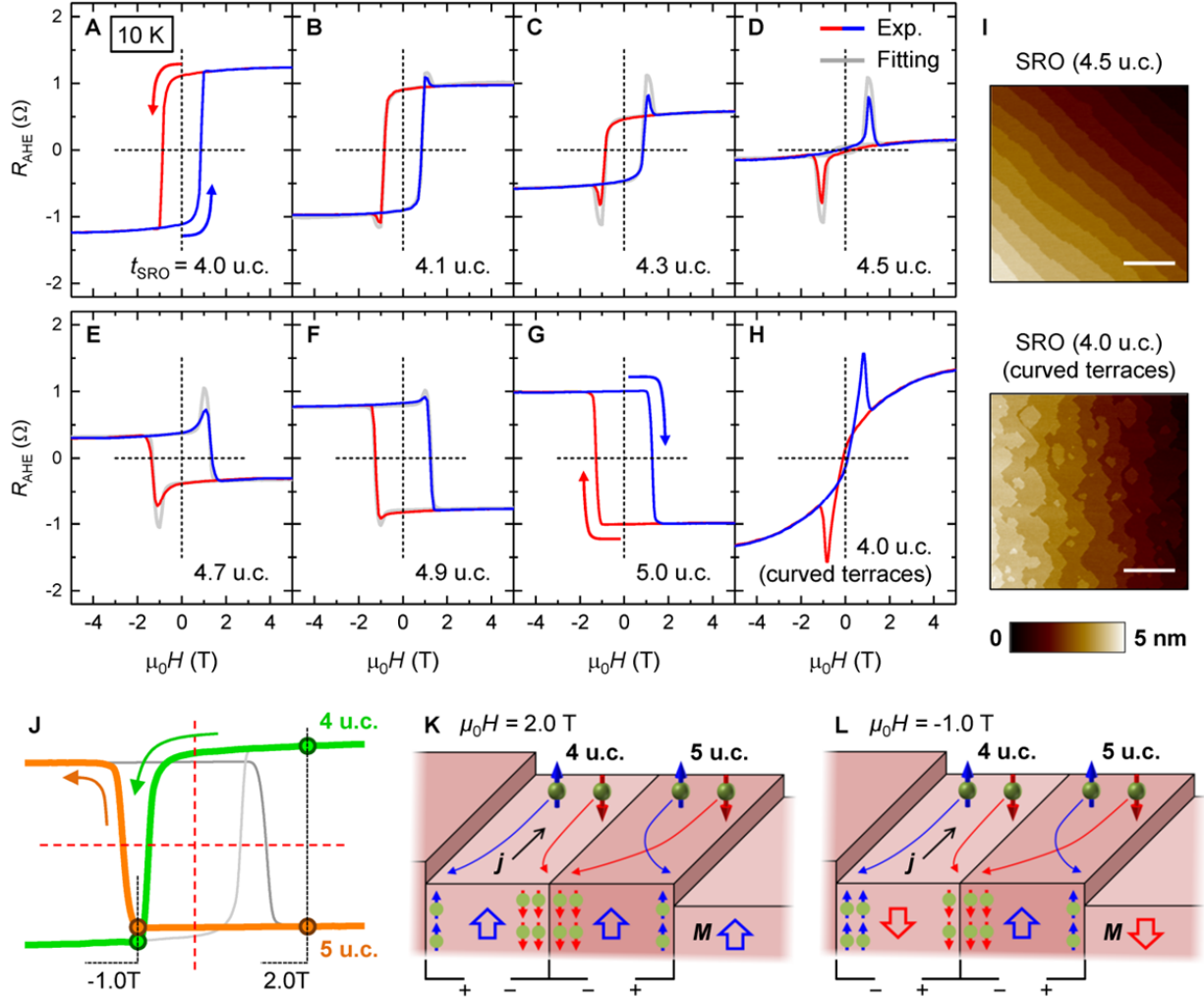

**Fig. 4. Two-channel anomalous Hall effect of SRO films with thickness inhomogeneity.** (**A** to **G**) $R_{AHE}$–$H$ curves of the SRO films with $t_{SRO}$ ranging from 4.0 to 5.0 u.c. $R_{AHE}$–$H$ curves of SRO samples with nominal $t_{SRO}$ = 4.1 ~ 4.9 u.c. show hump-like features similar to THE. The fitting curves (light grey) with simple linear superpositions of $R_{AHE}$–$H$ curves from 4.0 and 5.0 u.c. SRO films are also inserted for comparison. (**H**) $R_{AHE}$–$H$ curve of an SRO film grown under the high laser repetition rate of 10 Hz, which shows much more significant hump features. (**I**) Atomic force microscopy (AFM) images from the 4.5 u.c. SRO film grown under optimal condition (upper panel) and the 4.0 u.c. SRO film grown under high laser repetition rate (bottom panel). (**J** to **L**) Schematic illustrations of two-channel AHE in the 4.5 u.c. SRO film as $H$ scans from positive to negative. When $M$ of 4.0 and 5.0 u.c. channels align parallel (at 2.0 T), electrons in the majority bands accumulate at the opposite edges (**K**), making the $R_{AHE}$ signals in the two channels cancel out (**J**). On the contrary, when $M$ of the 4.0 and 5.0 u.c. channels align antiparallel (at -1.0 T), $R_{AHE}$ in the two channels share the same sign and contribute cumulatively to the overall $R_{AHE}$ (**J** and **L**).

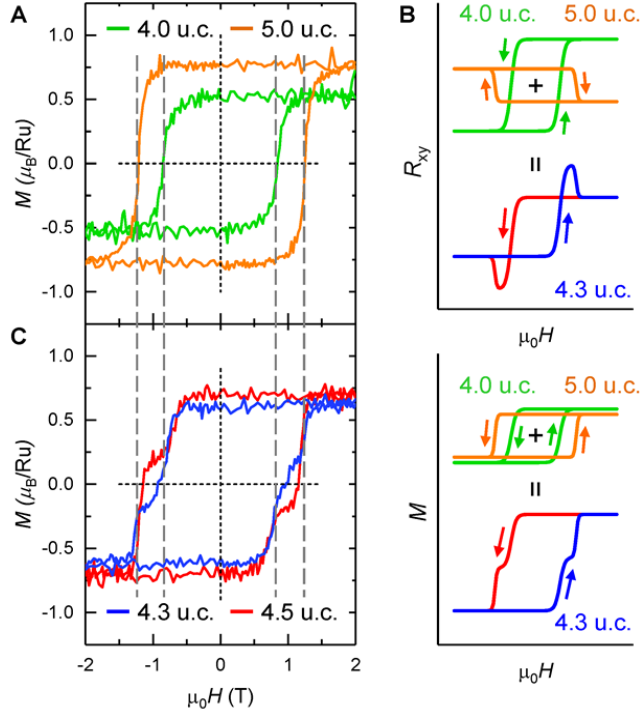

**Fig. 5. Two-step magnetic switching in the SRO films with inhomogeneous $t_{SRO}$.** (**A**) $H$-dependent magnetization ($M$–$H$) curves measured from the 4.0 and 5.0 u.c. SRO films at 10 K. (**B**) Schematic $R_{AHE}$–$H$ and $M$–$H$ curves of the SRO film with nominal $t_{SRO}$ = 4.3 u.c., which can be seen as a simple superposition of the curves of 5.0 and 4.0 u.c. films. This 4.3 u.c. SRO film should simultaneously exhibit hump-like feature in $R_{AHE}$–$H$ curve and two-step features in the $M$–$H$ curve. (**C**) $M$–$H$ curves measured from the 4.3 and 4.5 u.c. SRO films at 10 K. The $M$–$H$ curves clearly exhibit step-like features, and the two $H_C$ show good correspondence with the $H_C$ values of 4.0 and 5.0 u.c. SRO films.

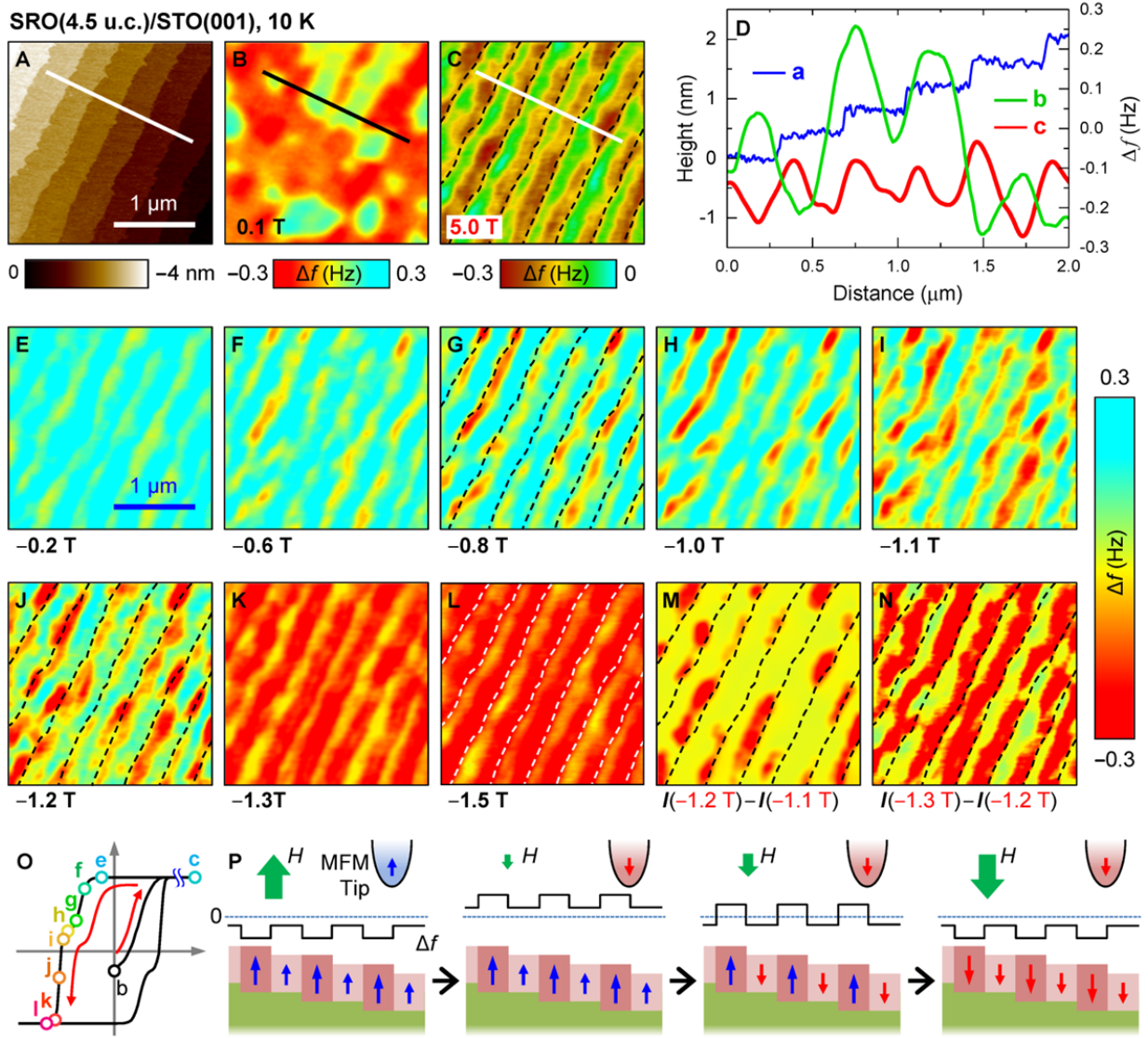

**Fig. 6. Magnetic force microscopy on the ferromagnetic domain structure and switching behavior of the 4.5 u.c. SRO film.** (**A**) AFM topographic image measured from a 4.5 u.c. SRO film in a preselected region. (**B**) Magnetic force microscopy (MFM) image acquired after zero-field cooling the film to 10 K. (**C**) MFM image measured at 5.0 T. The scale of the resonant frequency shift ($\Delta f$) is zoomed to 0 ~ -0.3 Hz, to highlight the magnetic inhomogeneity within each terrace. (**D**) Cross-sectional AFM and MFM line profiles corresponding to the solid lines marked in **A** to **C**. (**E** to **L**), MFM images acquired from -0.2 to -1.5 T. (**M**, **N**), MFM images obtained via pixel-by-pixel subtraction of the images shown in **I** to **K**. The dashed lines in **C**, **G**, **L**, **M**, and **N** indicate the terrace edges. The $\Delta f$ scale for **E** to **N** is fixed at 0.3 ~ -0.3 Hz. (**O**) Schematic $M$-$H$ loop of the 4.5 u.c. SRO film. $H$ values for acquiring MFM images are marked by open circles. (**P**) Schematics for describing the magnetic domain switching behaviors and corresponding MFM contrasts as $H$ scans from positive (pointing upward) to negative (pointing downward). Since 4.0 and 5.0 u.c. thick regions have different $H_C$, the 4.5 u.c. SRO film exhibits two-step magnetic switching behavior.

**Supplementary Materials for**

# Controllable thickness inhomogeneity and Berry-curvature-engineering of anomalous Hall effect in SrRuO$_3$ ultrathin films

Lingfei Wang[1,2]*, Qiyuan Feng,[3,4] Han Gyeol Lee,[1,2] Eun Kyo Ko[1,2], Qingyou Lu,[3,4]*
and Tae Won Noh[1,2]*

Correspondence and requests for materials should be addressed to:

L.F.W. (email: lingfei.wang@outlook.com), Q.Y.L (email: qxl@ustc.edu.cn), and T.W.N. (email: twnoh@snu.ac.kr).

**This file includes:**

**Fig. S1.** Longitudinal transport and magnetization of SrRuO$_3$ films with various thicknesses.
**Fig. S2.** Longitudinal transport of SrRuO$_3$ films with thickness inhomogeneity.
**Fig. S3.** Scanning transmission electron microscopy of a 15 u.c. SrRuO$_3$/SrTiO$_3$(001) film.
**Fig. S4.** Magnetic hysteresis loops of BaTiO$_3$/SrRuO$_3$ bilayers with magnetic skyrmions.

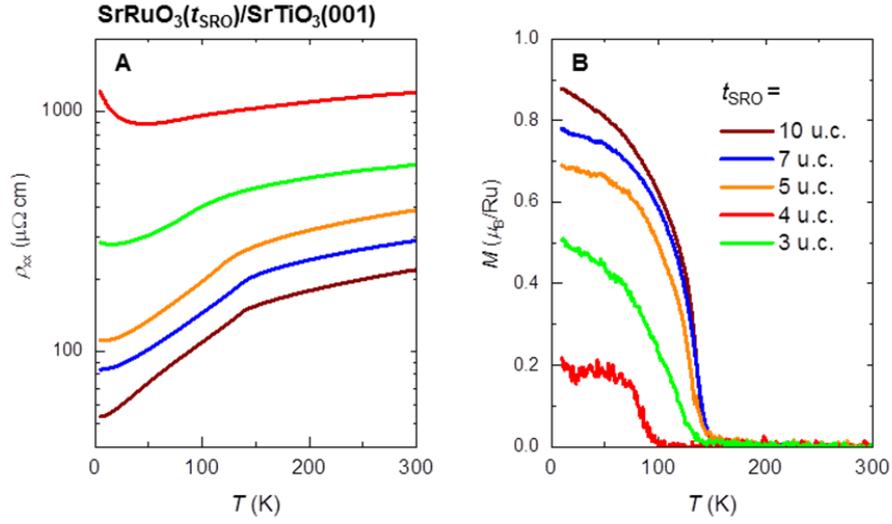

**Fig. S1. Longitudinal transport and magnetization of SrRuO$_3$ films with various thicknesses.** (**A**) Temperature-dependent longitudinal resistivity ($\rho_{xx}$–$T$) curves measured from SrRuO$_3$ (SRO) films with various thickness ($t_{SRO}$). The films were grown on SrTiO$_3$(001) [STO(001)] substrate. All the films were patterned into a 50×50 μm$^2$ Hall bar. (**B**) $T$-dependent magnetization ($M$–$T$) curves measured from SRO films with various $t_{SRO}$. Here the nominal $t_{SRO}$ values are all integers. The metallicity and ferromagnetism decay gradually as $t_{SRO}$ decreases.

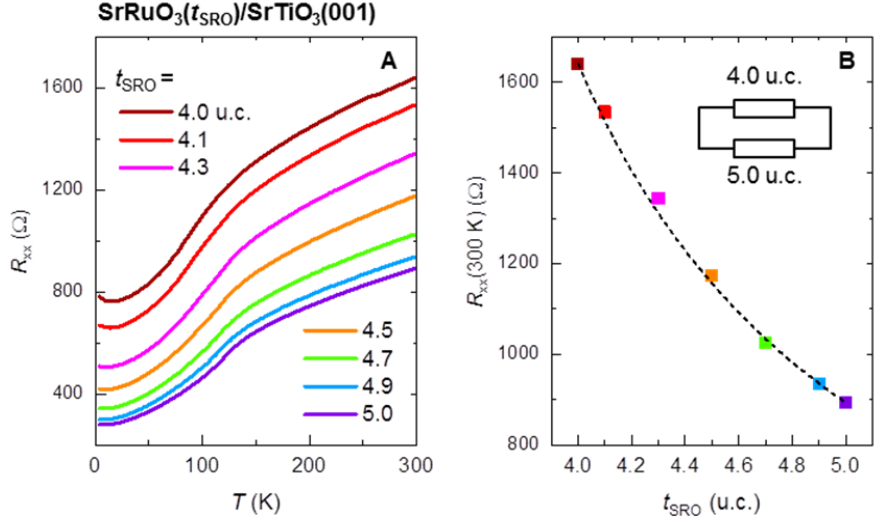

**Fig. S2. Longitudinal transport of SRO films with inhomogeneous $t_{SRO}$.** (**A**) $T$-dependent longitudinal resistance ($R_{xx}$–$T$) curves measured from SRO films with non-integer $t_{SRO}$ ranging from 4.1 to 4.9 u.c. All the films were patterned into a 50×50 μm$^2$ Hall bar. (**B**) $t_{SRO}$-dependent $R_{xx}$ at 300 K. The $R_{xx}$ versus $t_{SRO}$ curve can be well reproduced by a fitting curve which considers the resistance of 4.0 and 5.0 u.c.-thick regions are connected in parallel. This fitting result further attests our ability to achieve sub-unit-cell control of $t_{SRO}$.

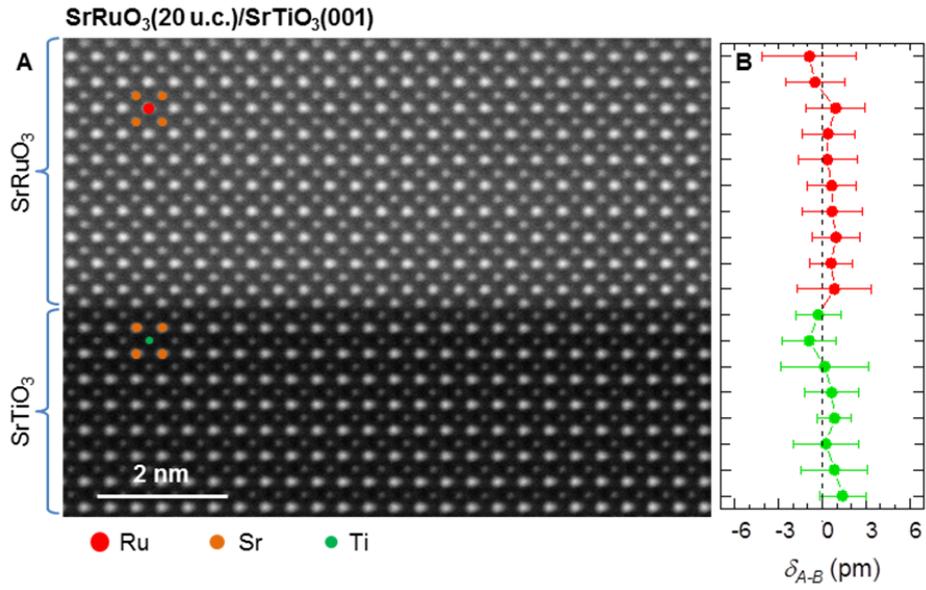

**Fig. S3. Scanning transmission electron microscopy at SRO/STO interface.** (**A**) Scanning transmission electron microscopy-High angle annular dark field (STEM-HAADF) image measured from 15 u.c. SRO film near the SRO/STO heterointerface. We choose to measure the 15 u.c.-thick film instead of a thinner one (e.g., 5 u.c.) to protect the interface structure from radiation damages during STEM specimen preparation. (**B**) Ferroelectric-like lattice distortion ($\delta_{A-B}$) near the SRO/STO interface. $\delta_{A-B}$ values were calculated from the displacements of B-site cations with respect to the centers of four A-site cations were calculated in each perovskite unit cell. In both SRO film and STO(001) substrate, $\delta_{A-B}$ is close to zero, which suggest that the Dzyaloshinskii -Moriya interaction in this SRO/STO(001) films are negligible.

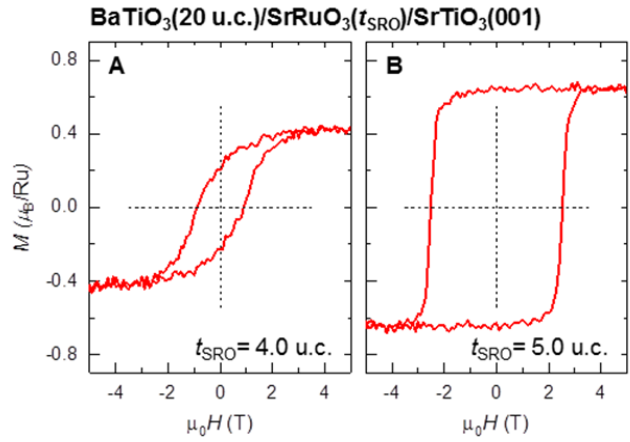

**Fig. S4. Magnetic hysteresis loops of BaTiO$_3$/SRO bilayers.** *M–H* curves of the BaTiO$_3$(20 u.c.)/SRO bilayers with $t_{SRO}$ = 4.0 u.c. (**A**) and 5.0 u.c. (**B**). These bilayer samples show magnetic skyrmions and topological Hall effect [L. Wang *et al.*, *Nat. Mater.* **17**, 1087–1094 (2018)], but the *M–H* curves do not exhibit any two-step features.